\documentclass[runningheads]{llncs}

\usepackage[T1]{fontenc}
\usepackage[utf8]{inputenc}
\usepackage[english]{babel}
\usepackage{microtype} 
\usepackage{amsmath,amssymb}
\usepackage{graphicx}
\usepackage{tikz}
\usetikzlibrary{arrows.meta,positioning,shapes.geometric,fit,backgrounds,calc}
\usepackage{booktabs}
\usepackage{multirow}
\usepackage{tabularx}
\usepackage{array}
\usepackage{adjustbox}
\usepackage{rotating}
\usepackage{xcolor}
\usepackage{algorithm}
\usepackage[noend]{algpseudocode}
\usepackage{cite}
\usepackage{url}
\usepackage{pifont}
\usepackage{hyperref}
\usepackage[nobiblatex]{xurl}

\hypersetup{bookmarksdepth=3, bookmarksopen=true, bookmarksopenlevel=1}

\graphicspath{{figures/}{../figures/}}

\usepackage{todonotes}
\usepackage{comment}

\begin{document}

\title{A Security Meta-Model for Retrieval-Augmented Generation Systems}

\ifdefined\anonymousversion\else\def\limosversion{1}\fi

\ifdefined\limosversion
\author{Steve Nouyep \and
S\'ebastien Salva \and
Maxime Puys}
\institute{Université Clermont Auvergne, CNRS, LIMOS, 63000 Clermont-Ferrand, France\\
\email{\{steve\_durand.nouyep\_noundjio, sebastien.salva, maxime.puys\}@uca.fr}}
\fi

\maketitle

\begin{abstract}

Retrieval-Augmented Generation (RAG) systems extend large language models
(LLMs) with external knowledge through a multi-stage pipeline. While this
architecture can improve the factual grounding of generated answers, it
introduces structural attack surfaces that extend beyond those of
standalone LLMs.
In this paper, we introduce a security meta-model that captures explicit
causal relationships between RAG surfaces, attacks, weaknesses, risks,
and CIA impact (Confidentiality, Integrity, Availability).
Its purpose is to provide security engineers with a structured and
user-friendly framework for gathering and assessing the risks,
weaknesses, and mitigations relevant to their RAG deployment.
We designed the meta-model through an iterative, structured analysis of
43~publications (2023--2026) and instantiated it as a catalog populated
with the security threats and remediations reported in the literature.
Filtering the catalog according to a deployment configuration produces
a risk profile containing the risks applicable to that deployment.
An interactive web visualizer lets users navigate the catalog as a graph, 
follow causal chains, and explore stakeholder-specific views.
Analysis of the catalog revealed a persistent imbalance between
attack-focused and defense-focused research, a concentration of
threats at ingestion, and coverage gaps affecting output
integrity.
Coverage is assessed against the OWASP LLM Top~10,
and operational applicability is illustrated across textual,
graph-based, and multimodal RAG configurations.

\keywords{RAG \and Threat Modeling \and Meta-model \and Risk Filtering}
\end{abstract}

\section{Introduction}
\label{sec:intro}

Retrieval-Augmented Generation (RAG)
systems~\cite{lewis2020rag} have emerged as a widely adopted approach for enhancing Large Language Models (LLMs) with
external knowledge sources.
By combining retrieval and generation, they aim to improve
factual grounding and domain adaptation in applications
ranging from enterprise knowledge management to healthcare
and e-commerce.
However, the architectural decomposition that enables these
benefits also introduces new security considerations.
Unlike standalone LLMs, RAG systems expose diverse pipeline
stages (document ingestion, indexing, retrieval, and
generation), each of which may constitute a distinct attack
surface.

Several studies and surveys~\cite{owasp_llm_top10,arzanipour2025threat,ammann2025risk,wang2025rag_security,bodea2026sok_rag_privacy} examine these attacks, propose mitigation strategies, or discuss their associated risks. 
While these contributions provide valuable insights into RAG security threats, they often address individual aspects in isolation, contributing to a growing body of (often complex) documentation. 
As a result, it becomes increasingly difficult for security engineers to identify and select the most appropriate solution for a given context.


These solutions either address LLM application security generically (e.g., OWASP LLM Top~10~\cite{owasp_llm_top10}) or analyze selected RAG threat classes; what remains less explored is a structured modeling approach that (i)~systematically links pipeline components to weaknesses, attacks, risks, and Confidentiality, Integrity, Availability (CIA) impact, and (ii)~adapts the assessment to concrete deployment configurations.

To help security engineers design secure RAG systems, it is essential to provide a structured and coherent view linking together RAG system types, attacks, weaknesses, risks, and mitigation strategies, while ensuring that this perspective remains adaptable to heterogeneous deployment contexts.

\smallskip\noindent\textbf{Contributions.}
To address this need, we propose in this paper:
\begin{enumerate}
\item \textbf{A security meta-model} that specifies entities
  and relations capturing structural relationships among RAG
  components, attacks, weaknesses, risks, and CIA impact
  (Section~\ref{sec:metamodel}), instantiated as a security
  catalog covering 3 RAG types, 8 attack surfaces,
  38 attacks, 12 weaknesses, 41 risks, and 72 mitigations. It is publicly
  available at \url{https://anonymous.4open.science/r/rag-mmsecu/}.
\item \textbf{A context-dependent filtering mechanism} that,
  given a RAG deployment configuration (e.g., network exposure,
  data source type), identifies the subset of applicable risks
  and produces a configuration-specific risk profile,
  illustrated on a multimodal e-commerce
  deployment~(Section~\ref{sec:filtering}).
\item \textbf{Four complementary taxonomic views} supporting
  distinct stakeholder roles (architect, CISO, pentester,
  DPO)~(Section~\ref{sec:views}).
\item \textbf{An interactive web visualizer} that
  operationalizes the meta-model: it renders the catalog as a
  navigable graph, lets users follow causal chains, applies the
  deployment filtering interactively, projects the four
  stakeholder views, and validates newly drafted attacks
  against the integration rules.
\end{enumerate}

\begin{figure}[t]
\centering
\includegraphics[width=.9\textwidth]{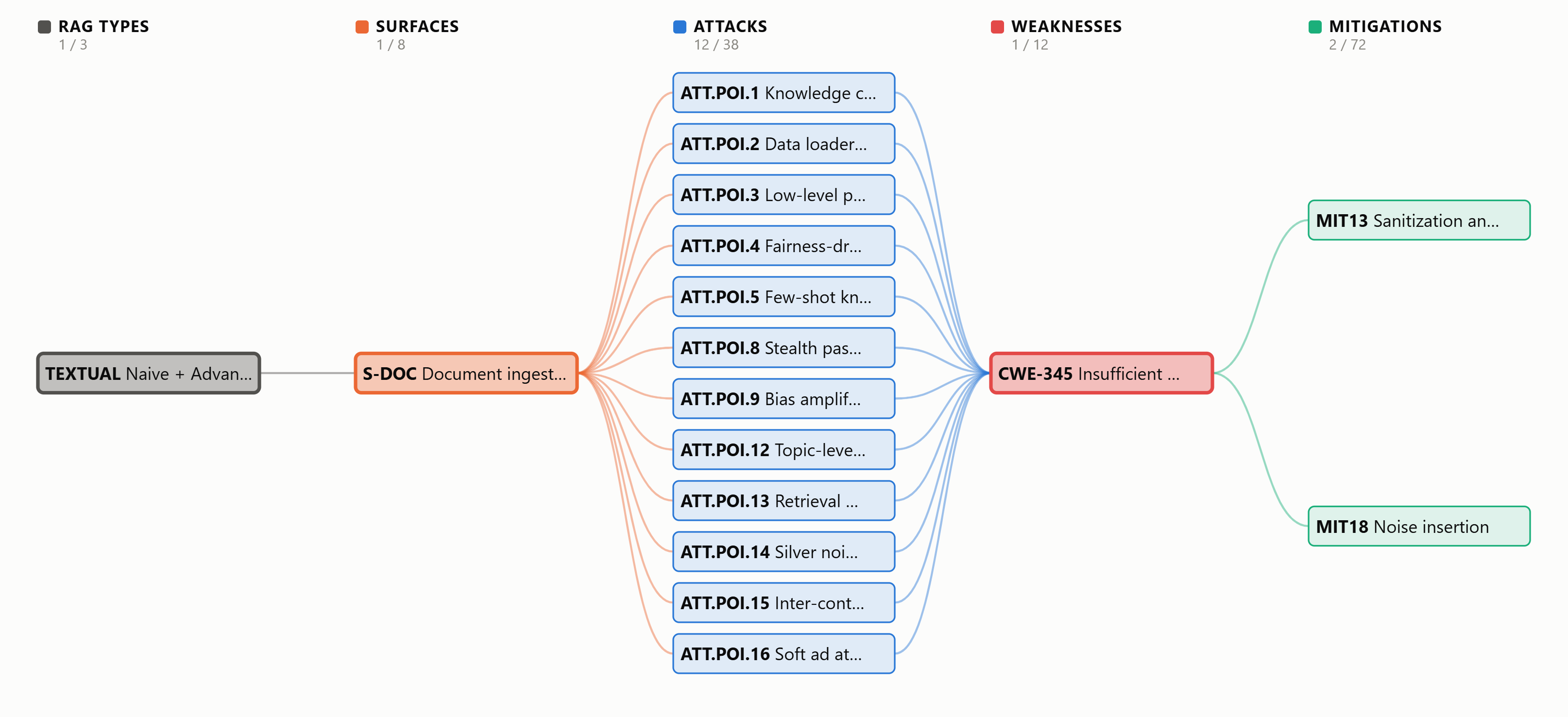}
\caption{The web visualizer rendering
the security catalog as a causal chain, here focused on
document-ingestion attacks: RAG type $\rightarrow$ surface
$\rightarrow$ attacks $\rightarrow$ exploited weakness
$\rightarrow$ mitigations. Per-column counters show the
narrowing effect of the projection (e.g., 12 of 38~attacks).
Filtering and views are detailed in
Sections~\ref{sec:filtering} and~\ref{sec:views}.}
\label{fig:v1_partial}
\end{figure}

We designed the meta-model from the entity and relation types extracted through a structured analysis of 43~publications (2023--2026, plus the foundational RAG paper from 2020~\cite{lewis2020rag}). We then instantiated it as a catalog populated with the security data collected from these publications, and derived the four stakeholder views from this catalog.
This structured integration also revealed cross-cutting patterns, including a systematic imbalance between attack and defense research~(Section~\ref{sec:discussion}).
%

\textbf{Outline.} The remainder of the paper is structured as follows. Section~\ref{sec:background} presents background notions and reviews related work. Section~\ref{sec:methodology} details the corpus construction. Section~\ref{sec:metamodel} introduces the meta-model, its iterative derivation, and its instantiation into the security catalog. Section~\ref{sec:views} presents context- and role-related views of the catalog, exposed through an interactive web visualizer. Section~\ref{sec:discussion} discusses quality assessment, cross-cutting observations, a research agenda, limitations, and threats to validity. Section~\ref{sec:conclusion} concludes and outlines future directions.

\section{Background and Related Work}
\label{sec:background}


\subsection{RAG System Architecture and Analysis}
\label{sec:architecture}

RAG systems combine LLMs with an external retrieval mechanism to ground responses in external corpora~\cite{lewis2020rag}.

\begin{figure}[t]
\centering
\includegraphics[width=\textwidth]{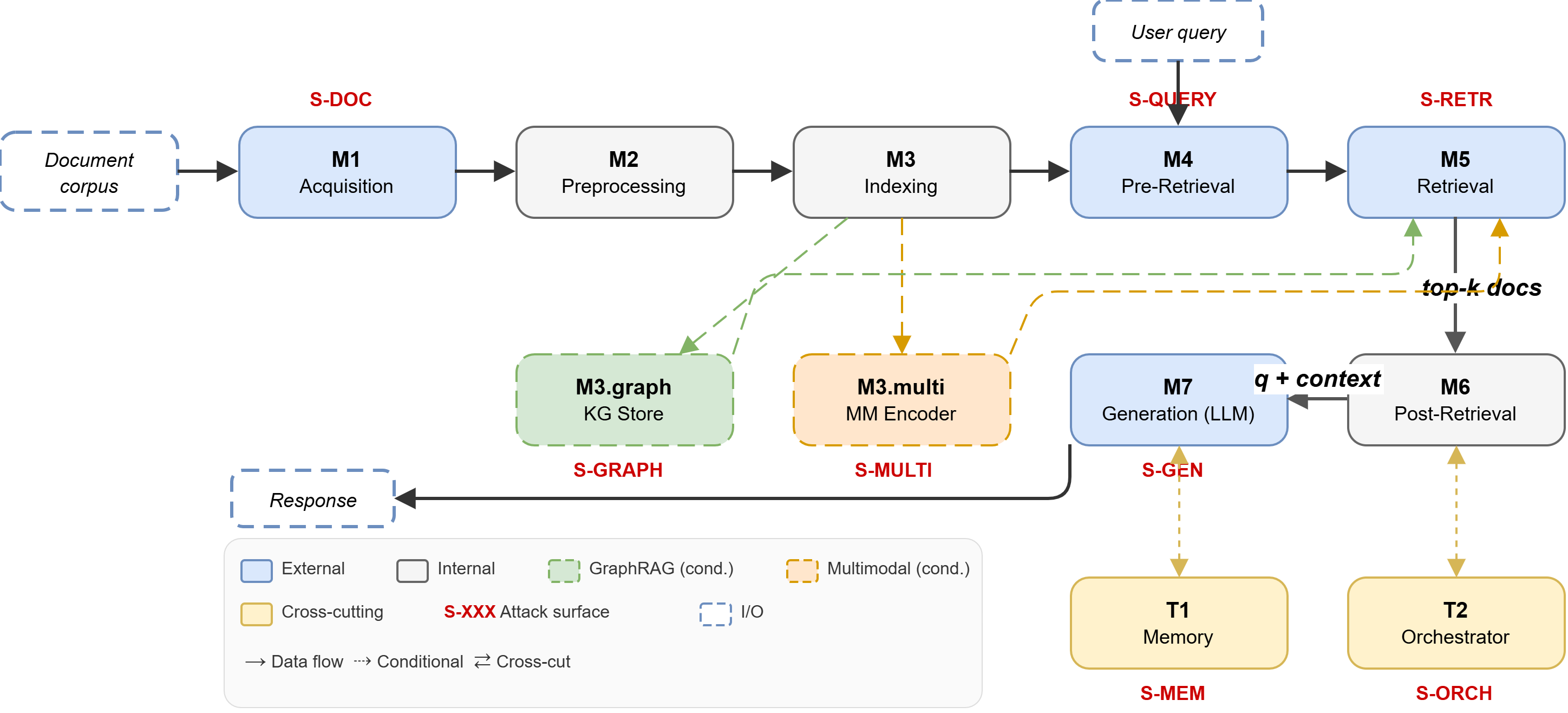}
\caption{RAG pipeline: 11~modules (M1--M7 sequential,
M3.graph/M3.multi conditional, two cross-cutting),
8~attack surfaces, and data flow.
Blue = external-facing; gray = internal; green/orange = conditional
(GraphRAG/Multimodal); yellow = cross-cutting; white dashed = I/O.}
\label{fig:pipeline}
\end{figure}

\smallskip\noindent\textbf{Module decomposition.}
Gao et~al.~\cite{gao2024rag_survey,gao2024modular_rag}
identify three paradigms (Naive, Advanced, Modular) and,
in their Modular RAG framework, define six module types
(Indexing, Pre-/Post-retrieval, Retrieval, Generation,
Orchestration) with 14~sub-modules and 40+~operators.
We derive from this framework a security-oriented
decomposition of 11~modules
(Figure~\ref{fig:pipeline}) by applying two refinements:

\begin{enumerate}
\item \textbf{Trust-boundary split of Indexing.}
  Gao et~al.'s Indexing type is split into three modules:
  document acquisition~(M1), preprocessing~(M2), and vector
  indexing~(M3).
  From a functional standpoint they form a single stage, but
  from a security standpoint M1 \emph{straddles} the external trust
  boundary (attacker-supplied documents), whereas M2 and M3
  are internal and expose no direct attack surface.
  Two conditional sub-modules extend M3: M3.graph (knowledge-graph
  indexing, active when type=GRAPHRAG) and M3.multi (multimodal
  encoding, active when type=MULTIMODAL).

\item \textbf{Cross-cutting modules.}
  Two modules operate orthogonally to the sequential pipeline:
  \emph{conversational memory} (which Gao et~al.\ identify but
  do not include in their catalog) and \emph{agentic
  orchestration} (corresponding to Gao et~al.'s Orchestration
  type).
  Both are optional, activated by deployment properties
  (Section~\ref{sec:filtering}), and each exposes a dedicated
  attack surface when enabled.
\end{enumerate}

\noindent The resulting 11~modules expose 8~attack surfaces
(S-DOC, S-QUERY, S-RETR, S-GEN, S-GRAPH, S-MULTI, S-MEM,
S-ORCH); three internal modules (M2, M3, M6) expose no direct
surface but can be targeted indirectly via upstream entry
points~\cite{greshake2023injection,zou2025poisonedrag}.

\smallskip\noindent\textbf{RAG type grouping.}
While the three RAG paradigms cited previously (Naive, Advanced, Modular)  differ operationally, our analysis of
their operators~\cite{gao2024modular_rag} indicates that they expose
comparable attack surfaces and CIA impact patterns, when neither graph-based indexing nor multimodal encoders are employed.
We therefore group them under the umbrella term
\textbf{TEXTUAL}, distinguishing them from \textbf{GRAPHRAG}
and \textbf{MULTIMODAL}, which introduce additional components
and specific attack surfaces. 
This grouping reflects a risk-oriented abstraction rather than
an architectural claim of equivalence; it independently
converges with the modalities dimension of
Nikishina et~al.~\cite{nikishina2025rag_taxonomy}.

\smallskip\noindent\textbf{Trust boundaries.}
From a security standpoint, two trust boundaries are essential.
First, external documents entering M1 may be
attacker-controlled and persist in storage.
Second, retrieved passages (M5) are integrated into the LLM
prompt~(M7): any attacker-controlled content in the corpus
can therefore directly influence the model's output, which is
the structural basis for indirect prompt injection attacks.
These properties motivate the need for a security model
explicitly aligned with pipeline stages.

\subsection{Threat Modeling Foundations}
\label{sec:standards}

Modeling RAG security requires bridging
information-security risk management with software
weakness enumeration.
We adopt the ISO/IEC~27000 and 27005
terminology~\cite{iso27000,iso27005}, distinguishing
threats, vulnerabilities, assets, and risks.
In this context, risk is treated as an \emph{integrative
concept} that captures the combined effect of a threat
exploiting a vulnerability and producing an impact on an asset.
This integrative role motivates our decision to model risks
explicitly rather than treating attacks or weaknesses in
isolation.

CAPEC patterns~\cite{mitre_capec} and CWE
identifiers~\cite{mitre_cwe} are widely used classification
systems for attack patterns and software weaknesses, providing
a common vocabulary across security tools and analyses.
NIST~SP~800-30~\cite{nist_sp800_30} models risk assessment as
a causal chain from threat source to impact.
Both serve as structural anchors for the meta-model of
Section~\ref{sec:metamodel}.

Finally, the OWASP LLM Top~10~\cite{owasp_llm_top10} provides an
external benchmark for assessing coverage of RAG-relevant
categories, while targeting LLM applications more broadly
rather than RAG architectures specifically, and its entries
carry no standard weakness identifiers.
MITRE ATLAS~\cite{mitre_atlas} provides a complementary
taxonomy of adversarial ML tactics and techniques, but targets
ML systems more generally and does not model RAG pipeline
stages or deployment-dependent applicability.
Our attacks could similarly be mapped to ATLAS techniques, but
we leave this alignment for future work.

\subsection{RAG and LLM Security Frameworks}
\label{sec:related}

Existing work on RAG and LLM security can be grouped into
(i)~risk-oriented frameworks~\cite{ammann2025risk,owasp_llm_top10}, (ii)~formally defined threat
models~\cite{arzanipour2025threat}, and (iii)~descriptive surveys of attacks and defenses~\cite{wang2025rag_security,bodea2026sok_rag_privacy}.

The OWASP LLM Top~10 (2025)~\cite{owasp_llm_top10} provides a
practitioner-oriented, consensus-based ranking of ten risk
categories for LLM applications.
While widely adopted, it targets LLM applications generically:
RAG pipeline stages are not modeled, categories are not mapped
to CWE/CAPEC identifiers, and retrieval-related risks (e.g.,
data poisoning and sensitive information disclosure) appear as
subcases of broader categories rather than architecturally
distinct threats.

In 2025, Arzanipour et~al.~\cite{arzanipour2025threat} propose a
formal threat model for RAG systems.
They identify three confidentiality-oriented attack classes
(membership inference (DL-MIA), content extraction, and data
poisoning) and formalize them through game-based security
definitions with an adversary-challenger paradigm.
Their adversary taxonomy distinguishes four types along two
axes: API-level vs.\ knowledge-base access and black-box vs.\
informed.
The scope remains limited to confidentiality-focused threats;
broader integrity and availability threats are outside the
model.
Differential privacy is the primary formally analyzed defense,
while other mitigations are discussed conceptually.
The model provides no CWE/CAPEC mapping, RAG-variant
differentiation, or deployment awareness.

Ammann et~al.\ (2025)~\cite{ammann2025risk} present a layered
risk assessment framework identifying eleven RAG-specific risks
and thirteen mitigations in a binary risk-mitigation matrix.
It combines IT baseline, AI/LLM, and RAG-specific layers,
anchored in OWASP, NIST~RMF, and ISO~27001, and is demonstrated
on three use cases.
The approach is component-centric: risks map directly to
pipeline elements without typed intermediate entities (attacks,
weaknesses) or explicit causal chains.
It also provides no CWE/CAPEC-level mappings, RAG-type
distinction, or deployment-based risk filtering.

Wang et~al.\ (2025)~\cite{wang2025rag_security} survey RAG
security from a full-chain perspective, identifying four threat
classes (adversarial attacks, data poisoning, backdoor attacks,
and privacy leaks) and eight defense techniques organized into
three layers: input hardening, data protection, and privacy
preservation.
They identify the \emph{Phantom} trigger-based backdoor as an
emerging single-document threat.
However, the survey does not follow an explicit systematic
review protocol, relate threats to security properties, or
distinguish RAG pipeline stages; multimodal RAG is only
discussed as future work.

Bodea et~al.\ (2026)~\cite{bodea2026sok_rag_privacy} present the
most closely related work: a systematic literature review of
72~publications that develops a taxonomy of RAG privacy risks
(leakage and adversarial manipulation) and a mitigation catalog
scored by relevance and maturity.
Its scope is restricted to confidentiality, with integrity and
availability outside its boundaries.
It covers only naive textual RAG, with variant architectures
identified as an open gap, and does not provide CWE/CAPEC
mapping, deployment filtering, or explicit causal chains.

Table~\ref{tab:positioning} summarizes these frameworks using
ten criteria, covering scope (RAG specificity, number of
publications analyzed), modeling depth (explicit specification,
causal chain, CWE/CAPEC mapping), and operational support (RAG
types, deployment filtering, mitigations, taxonomic views).
Our work complements them in three ways.
First, it provides explicit structural modeling across RAG
pipeline stages using typed entities and relations (rows
\emph{Explicit specification} and \emph{Explicit causal chain});
the former is matched only by Arzanipour et~al., while the
latter is at best partial in the other frameworks.
Second, it provides traceability to CWE/CAPEC identifiers,
which none of the compared frameworks provides.
Third, it provides deployment-aware filtering that adapts the
risk profile to a given RAG configuration (row \emph{Deployment
filtering}, with six deployment properties).
It also covers all three CIA dimensions, the only entry in
Table~\ref{tab:positioning} to do so (row \emph{CIA coverage}).

\begin{table}[t]
\caption{Positioning relative to existing work.}
\label{tab:positioning}
\centering\scriptsize
\setlength{\tabcolsep}{2.5pt}
\begin{tabular}{@{}lcccccc@{}}
\toprule
\textbf{Criterion} & \textbf{OWASP} & \textbf{Arzan.} & \textbf{Amm.} & \textbf{Wang$^c$} & \textbf{Bodea} & \textbf{Ours} \\
\midrule
RAG-specific            & Partial & \checkmark & \checkmark & \checkmark & \checkmark & \checkmark \\
Explicit specification  & ---     & \checkmark & ---        & ---        & ---        & \checkmark \\
Provable security       & ---     & \checkmark & ---        & ---        & ---        & --- \\
Explicit causal chain   & ---     & Partial    & Partial    & ---        & ---        & \checkmark \\
CWE/CAPEC mapping       & ---     & ---        & ---        & ---        & ---        & \checkmark \\
RAG types               & ---     & ---        & ---        & ---        & ---        & \checkmark\,(3) \\
Deployment filtering    & ---     & ---        & ---        & ---        & ---        & \checkmark\,(6) \\
Mitigations             & \checkmark & DP$^a$ & \checkmark\,(13) & \checkmark\,(8) & \checkmark & \checkmark\,(72) \\
Taxonomic views         & ---     & ---        & ---        & ---        & ---        & \checkmark\,(4) \\
Publications analyzed   & Ind.$^b$& $\sim$15   & $\sim$20   & 25         & 72         & 43 \\
\midrule
Methodology             & Consensus & Formal   & Risk assess. & Narr.\ survey & SoK & Iter.\ taxonomy \\
CIA coverage            & C,I       & C,I      & C,I          & C,I           & C   & C,I,A \\
\bottomrule
\multicolumn{7}{@{}l@{}}{\scriptsize $^a$Defensive-posture recommendations only (no specific countermeasures).} \\
\multicolumn{7}{@{}l@{}}{\scriptsize $^b$Community consensus, not a systematic literature corpus.} \\
\multicolumn{7}{@{}l@{}}{\scriptsize $^c$Wang et~al.~\cite{wang2025rag_security} (RAG security survey); distinct from~\cite{wang2025sok_supply_chain,wang2026silent_leaks}.} \\
\end{tabular}
\end{table}

\subsection{Broader Systematizations and Empirical Studies}

Several recent systematizations address related but broader
scopes.
Wang et~al.~\cite{wang2025sok_supply_chain} empirically analyze
529~vulnerabilities across 77~LLM projects, but treat RAG as
one application-layer component among many.
Duddu et~al.~\cite{duddu2024sok_interactions} show that
combining ML defenses may degrade security, a principle
relevant to RAG but studied only on classifiers.
Schroer et~al.~\cite{schroer2025sok_offensive} examine offensive
AI capabilities, while Hong et~al.~\cite{hong2025sok_prompt}
categorize prompt security for standalone LLMs.
None provides a unified structural model of RAG-specific threats
across the three CIA dimensions.

\smallskip\noindent\textbf{Empirical evidence.}
Recent experiments converge on three findings that motivate a
structured security model.
First, retrieval reshapes rather than reduces risk: RAG
increases unsafe outputs in 8 of 11~models, even from safe
documents~\cite{an2025rag_not_safer}, and shifts exposure from
training data to retrieval data~\cite{zeng2024privacy_rag}.
Second, both integrity and confidentiality are practically
exploitable: knowledge-base poisoning achieves high success
rates under black-box access~\cite{zou2025poisonedrag}, while
implicit knowledge extraction succeeds through benign-looking
queries~\cite{wang2026silent_leaks}.
Benchmarking~\cite{liang2025saferag} and certified
bounds~\cite{kang2024crag} further support the practical
relevance of these threats.
Third, contextual factors such as access model, underlying LLM,
and data provenance affect severity, but their interplay has
not been systematically studied.

\section{Corpus Construction and Scope}
\label{sec:methodology}
\label{sec:corpus}

We conducted a structured literature analysis targeting
RAG-specific security publications.
The objective was structural representativeness, that is,
identifying sufficient patterns to derive a coherent meta-model
rather than the exhaustiveness of a formal systematic
literature review.

\smallskip\noindent\textbf{Search strategy.}
We searched three sources: Google~Scholar, IEEE~Xplore, and
arXiv.
We executed six keyword queries on arXiv (title search,
Computer Science categories, cross-listed papers included)
over January~2020--February~2026:
\emph{``RAG attack,''} \emph{``RAG poisoning,''}
\emph{``RAG privacy,''} \emph{``RAG security,''}
\emph{``prompt injection AND RAG,''}
and \emph{``RAG membership inference.''}
These queries returned 67~results before deduplication
(31, 11, 10, 10, 3, and 2~respectively); all security-relevant
results date from 2024 onward.
Complementary searches on IEEE~Xplore and Google~Scholar
captured published work.
When the full text was inaccessible, the corresponding arXiv
preprint was used, giving 78 articles in total.

\smallskip\noindent\textbf{Selection process.}
After cross-query and cross-database deduplication, 50~unique
candidates remained.
Backward snowballing identified additional publications not
captured by keyword queries, including papers using compound
names such as SafeRAG, BadRAG, and BiasRAG, as well as
foundational work predating 2024.
Inclusion required at least one RAG-specific attack or defense
with sufficient technical detail to identify the exploited
pipeline surface and resulting CIA impact.
Papers addressing LLM security without retrieval, or RAG
quality without a security dimension, were excluded.

\smallskip\noindent\textbf{Final corpus.}
The retained corpus comprises \textbf{43~publications}:
31~attack-focused~(72\%), 9~defense-focused~(21\%), and
3~frameworks~(7\%).
By venue, 14~are arXiv-only preprints~(33\%), 19~appear in
major conferences (AAAI, ACL, ACM~CCS, EMNLP, ICML, ICLR,
USENIX~Security, WWW; 44\%), and 10~in journals or
workshops~(23\%).
We mitigate the preprint share by cross-checking published
versions where available and applying the same inclusion
threshold to all publications.
Extracted entities are additionally grounded in CWE and CAPEC,
independent of publication maturity.
Normative references (ISO~27005, NIST~SP~800-30, CWE, OWASP)
provide further structural anchors.

This corpus serves two purposes in the remainder of the paper:
deriving the structure of the meta-model
(Section~\ref{sec:construction}) and populating its catalog
(Section~\ref{sec:integration}).


\section{The RAG Security Meta-Model}
\label{sec:metamodel}

We use the term \emph{meta-model} in the sense of B\'ezivin~\cite{bezivin2005models}, that is, a model defining the structure (entities, relations, constraints) in which concrete models are instantiated.
Our meta-model does not describe \emph{one} specific RAG system, but the common structure underlying security models of RAG systems, based on the literature review in Section~\ref{sec:corpus}.
It is extensible, as new attacks can be added without changing the structure, and actionable, as a RAG configuration automatically produces a configuration-specific risk profile.

We deliberately specify the meta-model using an entity--relationship style rather than heavyweight metamodeling standards (MOF/EMF) or OWL ontologies: ER diagrams and property graphs are familiar to security practitioners, and the integration rules that populate the model (Section~\ref{sec:integration}) require only boolean criteria over typed links.
This choice does not prevent a future OWL formalization if automated inference becomes necessary.

\smallskip
\noindent\textbf{Structural specification.}
Following B\'ezivin~\cite{bezivin2005models}, we define the
meta-model as a triple $\mathcal{M} = (E, R, C)$, where
$E$ is the set of entity types (RAG, Surface, Attack,
Weakness, Risk, Mitigation, CIA),
$R$ is the set of directed relation types linking them
(e.g., \emph{exposes}, \emph{enables}, \emph{exploits},
\emph{generates}, \emph{affects}, \emph{mitigates},
\emph{remediates}),
and $C$ is the set of structural constraints that valid
instances must satisfy:

\begin{itemize}
\item Every Attack must be enabled by at least one
Surface.
\item Every Attack must exploit at least one
Weakness.
\item Every Attack must generate at least one
Risk.
\item Every Risk must affect at least one
CIA dimension (Confidentiality, Integrity, or Availability).
\end{itemize}

Together, these constraints ensure that attacks, weaknesses,
and risks are structurally connected to the architecture and
its security impact.

\begin{figure}[t]
\centering
\includegraphics[width=\textwidth]{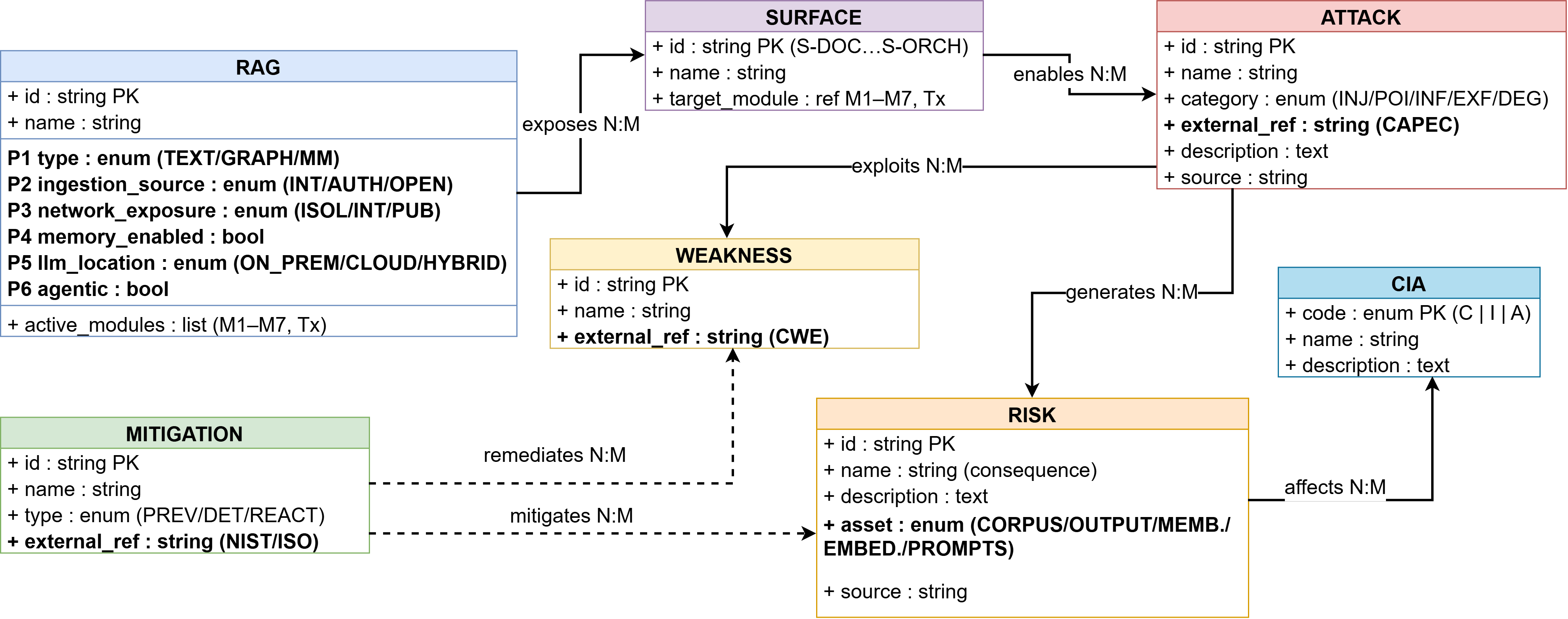}
\caption{Entity-relationship model: attributes, types,
and N:M cardinalities. Bold = type-discriminant;
P1--P6 drive risk filtering~(Section~\ref{sec:filtering}).}
\label{fig:er}
\end{figure}

\subsection{Iterative Construction of the Meta-Model}
\label{sec:construction}

The meta-model structure was derived following the iterative taxonomy
method of Nickerson et~al.~\cite{nickerson2013taxonomy}.
We first defined the \emph{meta-characteristic} shared by all
model elements: \emph{linking each RAG architectural surface to
its CIA impact through a causal chain}.
An initial conceptual-to-empirical iteration seeded the entity
types from ISO~27005 risk terminology and Ammann et~al.'s
component-centric framework~\cite{ammann2025risk}; subsequent
empirical-to-conceptual iterations refined this skeleton against
the corpus.
This process led to splitting indexing into acquisition,
preprocessing, and indexing modules with distinct trust
boundaries; introducing conditional surfaces (S-GRAPH, S-MULTI,
S-MEM, S-ORCH) for type- and deployment-specific attacks; adding
the Mitigation entity and its
\emph{mitigates}/\emph{remediates} distinction; and deriving
deployment properties P1--P6 from recurring applicability
conditions.

We stopped when Nickerson et~al.'s objective ending conditions
were met: all corpus publications had been examined and the last
iteration introduced no structural additions.
Structural saturation was reached after the first
35~publications: the remaining eight confirmed existing
categories but introduced \emph{no new entity or relation
types}.
Even when restricted to the 29~peer-reviewed publications, the 
corpus still instantiates all seven entity and seven relation types, 
supporting the structural coverage of the resulting meta-model shown in Figure~\ref{fig:er}.

\subsection{RAG Types and Modules}
\label{sec:types}

Three RAG types are defined by a risk bifurcation criterion: a type is created if and only if it introduces risks absent from other types (Table~\ref{tab:types}; formal justification in~Section~\ref{sec:justification}).

\begin{table}[t]
\caption{RAG types and exclusive risks.}
\label{tab:types}
\centering\scriptsize
\begin{tabular}{@{}llll@{}}
\toprule
\textbf{Type} & \textbf{Description} & \textbf{Module} & \textbf{Exclusive risks} \\
\midrule
TEXTUAL    & Naive+Advanced+Modular & Base (M1--M7)        & --- \\
GRAPHRAG   & + Knowledge Graph      & M3.graph (S-GRAPH)   & R17 \\
MULTIMODAL & + Non-textual encoders & M3.multi (S-MULTI)   & R24, R25, R29, R39 \\
\bottomrule
\end{tabular}
\end{table}

\noindent\textbf{Justification for TEXTUAL grouping.}
The three operational variants of text-based RAG (Naive, Advanced, Modular~\cite{gao2024modular_rag}) expose identical attack surfaces and CIA impact patterns in our corpus when graph and multimodal components are absent.
We therefore group them as TEXTUAL, creating distinct types only when new risks emerge: knowledge-graph indexing introduces graph poisoning (M3.graph), and multimodal encoding introduces cross-modal attacks (M3.multi).
The complete module-to-surface mapping is available in the supplementary material (\url{https://anonymous.4open.science/r/rag-mmsecu/}).

\subsection{Causal chain}
\label{sec:causal_chain}
The seven entity types form a causal chain: a RAG system
\emph{exposes} surfaces, which \emph{enable} attacks; attacks
\emph{exploit} weaknesses and \emph{generate} risks, which
\emph{affect} CIA dimensions; mitigations \emph{mitigate} risks
and \emph{remediate} weaknesses.
Figure~\ref{fig:er} illustrates the attribute schema.
The causal-chain design follows
NIST~SP~800-30~\cite{nist_sp800_30}, which models risk
assessment as a chain from threat source to impact: the
meta-model makes this chain explicit by linking RAG pipeline
surfaces to attacks, weaknesses, and CIA impacts.
The \emph{mitigates} relation is contextual (which risk is
reduced), while \emph{remediates} is structural (which weakness
is eliminated); only preventive mitigations carry both
relations, detective and reactive mitigations carrying only
\emph{mitigates} (IR8, Section~\ref{sec:integration}).

\subsection{Security Data Integration and Catalog}
\label{sec:integration}

Eight integration rules (IR1--IR8, Table~\ref{tab:rules})
specify how links between entities are constructed when
populating the meta-model.
They are organized into three groups, each with a distinct
objective and verification criteria.

\smallskip\noindent\emph{Group~1: Anchoring attacks in the
architecture} (IR1--IR3).
\textbf{Objective:} determine which surfaces are exposed by a
given RAG configuration, which attacks they enable, and which
weaknesses those attacks exploit.
IR1 requires the corresponding module to be active and the
surface accessible; IR2 requires an entry vector documented in
at least one publication; IR3 requires a traceable
CAPEC\,$\to$\,CWE mapping or an explicit root-cause analysis.
This mapping ensures interoperability with established
security taxonomies; it is not strictly one-to-one, as some
ML-specific integrity issues are only coarsely represented in
CWE and some RAG-specific attacks are only partially covered.
A link is created only when its criterion is satisfied;
otherwise, the entity remains in the catalog but unconnected.

\smallskip\noindent\emph{Group~2: Deriving consequences}
(IR4--IR6).
\textbf{Objective:} derive the risks resulting from each
attack, assign the affected asset, and determine the CIA
impact.
IR4 produces one primary risk per attack, corresponding to the
documented direct consequence, plus zero or more secondary
risks for reported cascading effects.
IR5 assigns the affected asset as a Risk attribute rather than
an entity, keeping the meta-model focused on the causal chain
while supporting the asset-based V4 view.
IR6 maps disclosure, modification, and denial to
Confidentiality, Integrity, and Availability, respectively,
following the standard information-security impact
classes~\cite{iso27005}.
Every attack must generate at least one risk, and every risk
must affect at least one CIA dimension.

\smallskip\noindent\emph{Group~3: Linking defenses}
(IR7--IR8).
\textbf{Objective:} connect mitigations to the risks they
reduce and the weaknesses they remediate.
IR7 requires evidence from one of five sources (attack-paper
defenses, CWE recommendations, defense papers, standards, or
logical deduction) that the mitigation can reduce likelihood
or impact.
IR8 links only \emph{preventive} mitigations (43/72) to
weaknesses, since detective and reactive measures address risks
without eliminating their underlying cause.
Each link carries a traceability tag; mitigations without
supporting evidence remain in the catalog but are not linked
to a risk.

\begin{table}[t]
\caption{Integration rules for instance construction.}
\label{tab:rules}
\centering\scriptsize
\setlength{\tabcolsep}{2.5pt}
\begin{tabular}{@{}c@{\,}cll>{\raggedright\arraybackslash}p{3.9cm}@{}}
\toprule
& \textbf{\#} & \textbf{Relation} & \textbf{Link} & \textbf{Criterion} \\
\midrule
\multirow{3}{*}{\rotatebox[origin=c]{90}{Anchoring}}
& IR1 & exposes     & RAG $\to$ Surf.       & Module active \emph{and} surface accessible \\
& IR2 & enables     & Surf.\ $\to$ Att.     & Entry vector in $\geq$1 publication \\
& IR3 & exploits    & Att.\ $\to$ Weak.     & CAPEC$\to$CWE mapping or root-cause \\
\midrule
\multirow{3}{*}{\rotatebox[origin=c]{90}{Conseq.}}
& IR4 & generates   & Att.\ $\to$ Risk      & 1~primary + 0..N secondary \\
& IR5 & assigns$^\star$     & Risk.\texttt{asset}   & Consequence heuristic \\
& IR6 & affects     & Risk $\to$ CIA        & Discl.\,$\to$\,C; modif.\,$\to$\,I; denial\,$\to$\,A \\
\midrule
\multirow{2}{*}{\rotatebox[origin=c]{90}{Def.}}
& IR7 & mitigates   & Mit.\ $\to$ Risk      & Reduces likelihood or impact$^{\star\star}$ \\
& IR8 & remediates  & Mit.\ $\to$ Weak.     & Preventive mitigations only (43/72) \\
\bottomrule
\multicolumn{5}{@{}l@{}}{\scriptsize $^\star$Attribute assignment, not an inter-entity relation (Asset $\notin E$).} \\
\multicolumn{5}{@{}l@{}}{\scriptsize $^{\star\star}$5~evidence sources: attack papers, CWE, defense papers, standards, deduction.} \\
\end{tabular}
\end{table}

\smallskip\noindent\textbf{The resulting catalog.}
\label{sec:catalog}
Applying these rules to the 43~publications of
Section~\ref{sec:corpus} produces the \emph{catalog}, the
instance of $\mathcal{M}$ containing all extracted entities
and their relations: 3~RAG categories, 8~surfaces, 38~attacks,
12~weaknesses, 41~risks, 72~mitigations, and the 3~CIA
dimensions.
The attacks fall into five categories:
Poisoning~(16), Exfiltration~(11), Inference~(5),
Injection~(4), and Degradation~(2); poisoning is the
largest~(42\%), followed by exfiltration~(29\%).

Each risk is characterized by an attack, an affected asset,
and one or more CIA dimensions, across five assets:
CORPUS~(20), OUTPUT~(12), MEMBERSHIP~(5), EMBEDDINGS~(3), and
PROMPTS~(1).
Integrity accounts for 20~risks, confidentiality for 17, and
availability for~1; three span multiple CIA dimensions, and
three (R0, R26, R28) capture emerging or organizational risk
scenarios anchored in normative sources rather than directly
derived from a cataloged attack.
The mitigations divide into 43 preventive, 20 detective, and
9 reactive, classified following ISO~27001 and NIST CSF.

The catalog includes 12~CWE mappings covering all five attack
categories: INJ maps to CWE-1427/74/693/863, POI to
CWE-345/94/1321, INF to CWE-200/203, EXF to CWE-200/212, and
DEG to CWE-400/770.
Seven of the eight surfaces are exercised by at least one
attack; S-ORCH is included for extensibility but has no
dedicated attack in the current corpus.
The catalog is stored as a property graph, and its structural
constraints are checked mechanically at import time.
Two derived artifacts must not be confused with the catalog:
filtering it with a specific RAG deployment configuration
(Section~\ref{sec:filtering}) produces a \emph{risk profile},
the subset of risks applicable to that deployment; and the
stakeholder \emph{views} of Section~\ref{sec:views} are
projections of the catalog, not new instances.

\smallskip\noindent
Table~\ref{tab:chains} gives four representative causal chains
covering the main attack categories (POI, INJ, EXF, and INF);
degradation attacks (DEG) follow the same structure but
primarily target availability.

\begin{table}[t]
\caption{Illustrative causal chains (one per major attack category).}
\label{tab:chains}
\centering\scriptsize
\setlength{\tabcolsep}{2.5pt}
\begin{tabular}{@{}lllllll@{}}
\toprule
\textbf{Surface} & \textbf{Attack} & \textbf{CWE} & \textbf{Risk} & \textbf{Asset} & \textbf{CIA} & \textbf{Mit.\,(\#)} \\
\midrule
S-DOC & Corpus poison. & 345 & R8  & CORPUS & I &
  MI3, MI14, MI17~(11) \\
S-GEN & Indirect inj.  & 1427 & R9 & OUTPUT & I &
  MI4, MI5, MI16~(7) \\
S-RETR & Data exf. & 200 & R2 & CORPUS & C &
  MI0, MI9, MI27~(15) \\
S-RETR & Memb.\ inf. & 200 & R4 & MEMB & C &
  MI0, MI1, MI62~(4) \\
\bottomrule
\multicolumn{7}{@{}l@{}}{\scriptsize Mit.\,(\#)=Mitigations (count). MEMB=MEMBERSHIP. C=Confidentiality, I=Integrity.} \\
\end{tabular}
\end{table}

\smallskip\noindent\textbf{Walkthrough.}
To illustrate how the integration rules apply sequentially to
construct an explicit causal chain, consider the first chain in
Table~\ref{tab:chains}:
IR1 determines that surface S-DOC is exposed (module~M1
active); IR2 links S-DOC to the corpus-poisoning attack
(entry vector documented in~\cite{zou2025poisonedrag});
IR3 maps that attack to CWE-345 via its CAPEC entry;
IR4 generates risk~R8 as the primary consequence (each
attack produces exactly one primary risk, plus zero or
more secondary risks when a single technique causes
cascading effects);
IR5 assigns asset CORPUS to~R8;
IR6 labels R8 as Integrity; and IR7 links the
11~mitigations whose evidence (from attack-paper
defenses, CWE recommendations, dedicated defense papers,
standards, or logical deduction) shows they reduce R8's
likelihood or impact.
The same sequence can be applied to any new attack without
changing the meta-model
structure~(Section~\ref{sec:filtering} details the deployment
properties that IR1 depends on).

\section{Views Related to Context and Role}
\label{sec:views}

The full catalog can be covered, but it is difficult to navigate as a whole, as most security questions concern only a subset of it. We therefore provide an interactive web visualizer (Section~\ref{sec:visualizer}) with purpose-built \emph{views}: context-related views (Section~\ref{sec:filtering}) filter the catalog by deployment configuration, while role-related views (Section~\ref{sec:roleviews}) present the elements relevant to a given stakeholder role.

\subsection{The Interactive Web Visualizer}
\label{sec:visualizer}
Both view families are exposed through an interactive web visualizer that renders the catalog as a navigable graph (Figure~\ref{fig:v1_partial}). From it, a user follows causal chains from surfaces to CIA impact, switches between the role-related views of Section~\ref{sec:roleviews}, applies the context-related filtering of Section~\ref{sec:filtering} with risks relabeled live as the deployment configuration changes, and drafts new attacks that are validated against the integration rules before they enter the catalog.
The visualizer, together with the full catalog and the datasets and scripts that generate it, is publicly available at \url{https://anonymous.4open.science/r/rag-mmsecu/}.

\subsection{Views Related to Context}
\label{sec:filtering}

Not every threat applies to every RAG configuration: an isolated
system has no network-based risks, while a text-only pipeline
has no multimodal threats.
We therefore complement the catalog with context-dependent
filtering, which dynamically annotates risks according to a
deployment configuration without modifying the meta-model
structure.

\smallskip\noindent\textbf{Deployment properties.}
A RAG deployment is characterized by six properties (P1--P6)
that determine risk applicability (Table~\ref{tab:properties}).
They were derived bottom-up from the 41~risks by grouping
deployment conditions with the same security implication.
For example, open Internet sources and untrusted third-party
suppliers are both captured by P2\,=\,OPEN.
A property was retained only if it produces at least one
non-NORMAL effect in Table~\ref{tab:filtering};
parameters affecting severity alone were excluded.

\begin{table}[t]
\caption{RAG deployment properties.}
\label{tab:properties}
\centering\scriptsize
\begin{tabular}{@{}clll@{}}
\toprule
\textbf{\#} & \textbf{Property} & \textbf{Values} & \textbf{Semantics} \\
\midrule
P1 & \texttt{type}              & TEXTUAL, GRAPHRAG, MULTIMODAL & Architecture \\
P2 & \texttt{ingestion\_source}  & INTERNAL, AUTHENTICATED, OPEN & Ingestion control \\
P3 & \texttt{network\_exposure}  & ISOLATED, INTERNAL, PUBLIC    & Network exposure \\
P4 & \texttt{memory\_enabled}    & true / false                  & Conversation history \\
P5 & \texttt{llm\_location}      & ON\_PREMISE, CLOUD, HYBRID    & LLM location \\
P6 & \texttt{agentic}           & true / false                  & Action execution \\
\bottomrule
\end{tabular}
\end{table}

\smallskip\noindent\textbf{Choice of risk as segmentation criterion.}
\label{sec:justification}
We considered weakness (CWE), surface, attack, and risk as
candidate criteria; Table~\ref{tab:discrim}
(Appendix~\ref{app:tables}) compares their discriminant power.
CWE has no type-specific items, while surfaces have only 2 of
8; risk and attack each have 5~type-specific items.
Risk is preferred because it carries strictly more information:
$\text{Risk} = f(\text{Attack},\allowbreak{}
\text{Weakness},\allowbreak{} \text{Asset},\allowbreak{}
\text{CIA})$.
A risk thus subsumes its attack, while the converse does not
hold.


\smallskip\noindent\textbf{Filtering algorithm.}
Each property value produces one of four effects:
ELIMINATES removes an impossible risk (e.g., graph-specific
risks in a text-only pipeline); MITIGATES reduces its
likelihood; AGGRAVATES increases its impact (e.g., a
cloud-hosted LLM increases confidentiality exposure); and ADDS
activates an additional attack surface (e.g., agentic
capabilities activate S-ORCH).
Each risk therefore receives one of four labels
(\textsc{eliminated}, \textsc{mitigated}, \textsc{aggravated},
or \textsc{normal}) that stakeholders can use to prioritize
analysis; the first three correspond directly to the effects
above.
The \textsc{adds} effect operates on surfaces rather than
risks: the risks of an added surface, if any, enter the profile
with the \textsc{normal} label unless another rule applies.
The rules in Table~\ref{tab:filtering} are processed
generically by Algorithm~\ref{alg:filtering}: new rules extend
the table without changing the procedure, and new deployment
properties can be added without altering the meta-model
schema.
The web visualizer implements this
filtering interactively, relabeling risks as the user changes
the deployment configuration.

\begin{table}[t]
\caption{Filtering table: property $\times$ value $\to$ effect.}
\label{tab:filtering}
\centering\scriptsize
\begin{tabular}{@{}llll@{}}
\toprule
\textbf{Property} & \textbf{Value} & \textbf{Effect} & \textbf{Risks} \\
\midrule
\texttt{type}              & $\neq$ GRAPHRAG   & ELIMINATES & R17 \\
\texttt{type}              & $\neq$ MULTIMODAL & ELIMINATES & R24, R25, R29, R39 \\
\texttt{ingestion\_source}  & $=$ INTERNAL      & MITIGATES  & R8, R11, R19, R20 \\
\texttt{network\_exposure}  & $=$ ISOLATED      & ELIMINATES & R22, R40 \\
\texttt{memory\_enabled}    & $=$ false         & ELIMINATES & R16 \\
\texttt{llm\_location}      & $\in$ \{CLOUD, HYBRID\} & AGGRAVATES & R2, R6, R7 \\
\texttt{agentic}           & $=$ true          & ADDS       & Surface S-ORCH \\
\bottomrule
\end{tabular}
\end{table}

\begin{algorithm}[t]
\caption{Context-dependent risk filtering.}
\label{alg:filtering}
\begin{algorithmic}[1]
\Function{ApplicableRisks}{$\mathit{config}$, $\mathit{Rules}$}
  \State $\mathit{label}[r] \gets \textsc{normal}$ \textbf{for each} risk $r$ in the catalog
  \ForAll{$(\mathit{prop},\, \mathit{op},\, \mathit{value},\, \mathit{effect},\, \mathit{targets})$ \textbf{in} $\mathit{Rules}$}
    \If{$\mathit{config}[\mathit{prop}] \;\mathit{op}\; \mathit{value}$}
      \If{$\mathit{effect} = \textsc{adds}$}
        \State \textbf{activate} surface $\mathit{targets}$ and its cataloged risks
      \Else
        \State $\mathit{label}[r] \gets \mathit{effect}$ \textbf{for each} $r \in \mathit{targets}$
      \EndIf
    \EndIf
  \EndFor
  \State \Return $\{\,(r, \mathit{label}[r]) : \mathit{label}[r] \neq \textsc{eliminated}\,\}$
\EndFunction
\end{algorithmic}
\end{algorithm}

\smallskip\noindent\textbf{Illustrative example.}
Consider a multimodal e-commerce search engine configured as: type\,=\,MULTIMODAL, ingestion\,=\,OPEN, network\,=\,PUBLIC, memory enabled, LLM\,=\,CLOUD, non-agentic.
Applying Algorithm~\ref{alg:filtering}, only R17 (graph-specific poisoning) is eliminated because the system does not use knowledge-graph indexing. No risks are contextually mitigated because ingestion is open, while three confidentiality risks (R2, R6, R7) are aggravated because the cloud-hosted LLM involves data processing by an external provider.
The result is \textbf{40~applicable risks out of~41}, the most exposed profile in our catalog. This illustrates that the primary value of filtering is not risk elimination but risk differentiation: even in this worst-case configuration, the algorithm distinguishes structurally impossible risks from aggravated ones, enabling practitioners to prioritize their analysis.

\subsection{Views Related to Role}
\label{sec:roleviews}

In practice, no single stakeholder requires the full catalog at once. Different stakeholders need distinct views, emphasizing elements and relations relevant to their role. For example, a security architect deciding which CWE weaknesses to remediate before deployment does not require detailed CIA-level impact data, whereas a data protection officer assessing GDPR exposure does not need the full surface-to-weakness dependency chain. We therefore define a \emph{view} as a projection of the catalog that retains only the entity types and relations relevant to a given role, producing a self-contained subgraph with reduced cognitive load.

Table~\ref{tab:views} summarizes four views, one per stakeholder role. Each row specifies which entity types and relations are \emph{projected} (visible in the subgraph) and which are \emph{hidden} (abstracted away). The projected elements define the subgraph that stakeholders navigate; hidden elements remain in the underlying catalog but are not required for the role’s primary question. We describe each view below.



\begin{table}[t]
\caption{Taxonomic views: projection per stakeholder role.}
\label{tab:views}
\centering\scriptsize
\begin{tabular}{@{}cllll@{}}
\toprule
\textbf{View} & \textbf{Audience} & \textbf{Question} & \textbf{Projected} & \textbf{Hidden} \\
\midrule
V1 & Architect & ``What weaknesses?'' & RAG, Surf., Att., Weak., Mit. & Risk, CIA \\
V2 & CISO      & ``What to do?''      & Att., Weak., Risk, Mit. & RAG, Surf., CIA \\
V3 & Pentester & ``Full chain''       & RAG, Surf., Att., Weak., Risk, Mit. & CIA \\
V4 & DPO       & ``What risks?''      & Risk, CIA, Mit.\,(+\,asset) & RAG, Surf., Att., Weak. \\
\bottomrule
\end{tabular}
\end{table}

\medskip
\noindent\textbf{V1-Architect (5 entities, 4 relations).}
This view focuses on the architectural perspective by projecting the chain
RAG $\to$ Surface $\to$ Attack $\to$ Weakness $\leftarrow$ Mitigation
through the relations \emph{exposes}, \emph{enables}, \emph{exploits},
and \emph{remediates}.
Risk and CIA are hidden, allowing the architect to reason directly
about how architectural components expose attack surfaces that lead
to weaknesses and their available remediations.
\emph{Question: where are the architectural entry points that may lead to risks?}
By navigating the pipeline from system components to weaknesses,
this view helps identify structural security issues early in the design.
Figure~\ref{fig:v1_partial} (page~\pageref{fig:v1_partial})
shows a partial V1 projection for \textsc{Textual} RAG,
restricted to surface S-DOC and weakness CWE-345; the full V1
subgraph contains 78~nodes and 140~relationships.

\medskip

\noindent\textbf{V2-CISO (4 entities, 4 relations).}
This view focuses on the risk-mitigation perspective by projecting the mapping
Attack $\to$ Weakness $\to$ Risk $\leftarrow$ Mitigation
(with Mitigation also $\to$ Weakness)
through the relations \emph{exploits}, \emph{generates}, \emph{mitigates}, and \emph{remediates}.
RAG, Surface, and CIA are hidden, allowing the CISO to reason directly about
which defenses address which risks.
\emph{Question: which mitigations exist and where are the coverage gaps?}
Risks are grouped by category and linked to their available mitigations,
making missing protections immediately visible.
Five risks (R17, R33, R34, R35, R36) currently have no documented mitigation.

\medskip

\noindent\textbf{V3-Pentester (6 entities, 6 relations).}
This view combines V1 and V2 to expose the complete chain
RAG $\to$ Surface $\to$ Attack $\to$ Weakness $\to$ Risk,
while Mitigation connects both to Weakness (\emph{remediates})
and Risk (\emph{mitigates}).
Only CIA is hidden.
\emph{Question: how can a structured security assessment be planned?}
The pentester navigates the graph in both directions:
forward along the V1 path to identify potential entry points, and backward along the V2 path to verify deployed defenses.

\medskip

\noindent\textbf{V4-DPO (3 entities, 2 relations, + asset filter).}
This view projects Risk $\to$ CIA and Mitigation $\to$ Risk
through the relations \emph{affects} and \emph{mitigates}, filtered
by the target asset (CORPUS, OUTPUT, MEMBERSHIP, EMBEDDINGS, PROMPTS).
RAG, Surface, Attack, and Weakness are hidden, enabling the DPO
to reason directly in terms of data assets and CIA impact rather
than technical attack chains.
\emph{Question: for a given asset, which risks apply and what mitigations exist?}
CORPUS concentrates 49\% of all risks (20/41),
while OUTPUT exhibits the highest mitigation gap (25\%:
3 risks without mitigation out of 12),
making this view directly usable as input to GDPR Data Protection Impact Assessments.

The four views are complementary projections, not
prescriptive steps: each stakeholder selects the view
matching their role and obtains a self-contained subgraph.
Because a view is simply a projection defined by a subset
of entity types and relations, practitioners may define
additional views tailored to their own needs.

\section{Discussion}
\label{sec:discussion}
\label{sec:findings}

\subsection{Quality Assessment}
\label{sec:validation}

We examine the meta-model along six quality criteria,
drawn from taxonomy evaluation
literature~\cite{nickerson2013taxonomy,unterkalmsteiner2023taxonomy}:
\emph{coverage} (does the catalog span the domain?),
\emph{determinism} (do the rules produce a unique result?),
\emph{repeatability} (can the process be reproduced?),
\emph{unambiguity} (is each entity clearly defined?),
\emph{usefulness} (does it serve its stakeholders?),
and \emph{navigability} (can users find relevant information?).

\smallskip\noindent\textbf{Coverage against OWASP LLM Top~10.}
Table~\ref{tab:owasp_coverage} maps each OWASP LLM Top~10 (2025)
entry~\cite{owasp_llm_top10} to our meta-model entities.
Seven entries are fully covered; two entries
are partially covered, meaning that our catalog captures their RAG-specific manifestation (e.g., poisoning via document ingestion for Supply Chain, or the orchestration surface for Excessive Agency), but does not address their broader LLM-generic aspects (e.g., training-time attacks or plugin vulnerabilities); Unbounded Consumption falls outside our scope as it concerns LLM resources rather than RAG-specific threats.

\begin{table}[t]
\caption{Coverage of OWASP LLM Top~10 (2025) by the meta-model.}
\label{tab:owasp_coverage}
\centering\scriptsize
\setlength{\tabcolsep}{2.5pt}
\resizebox{\columnwidth}{!}{%
\begin{tabular}{@{}lllll@{}}
\toprule
\textbf{OWASP Entry} & \textbf{Title} & \textbf{Mapping} & \textbf{Risks} & \textbf{Cov.} \\
\midrule
Prompt Injection (01)       & INJ      & CWE-1427/94; S-GEN    & R9, R10       & Full \\
Sensitive Info Discl.\ (02) & EXF/INF  & CWE-200/212           & R2, R4--R7    & Full \\
Supply Chain (03)           & POI      & S-DOC; CWE-345        & R8, R11       & Part.$^\dag$ \\
Data/Model Poison.\ (04)    & POI      & CWE-345/1321; S-DOC   & R8, R11--R13, R15 & Full \\
Improper Output (05)        & POI/DEG  & CWE-345/770; S-DOC    & R34--R36      & Full \\
Excessive Agency (06)       & ---      & S-ORCH; P6=agentic    & ---$^\ddag$   & Part.$^\ddag$ \\
Prompt Leakage (07)         & EXF      & CWE-200; S-GEN        & R6            & Full \\
Vector/Embed.\ (08)         & POI/EXF  & S-RETR; CWE-345       & R8, R2        & Full \\
Misinformation (09)         & POI/INJ  & S-DOC/S-GEN           & R19, R34      & Full \\
Unbound.\ Consump.\ (10)    & ---      & (LLM resource mgmt.)  & ---           & OOS \\
\bottomrule
\multicolumn{5}{@{}l@{}}{\scriptsize $^\dag$Beyond doc.\ ingestion = out of scope.
$^\ddag$Covered via S-ORCH; no dedicated risk IDs. OOS = Out of Scope.} \\
\end{tabular}}
\end{table}

\smallskip\noindent\textbf{Determinism, Unambiguity, and Repeatability.}
Entity definitions are grounded in established standards and
taxonomies (ISO~27005 for risk, and CWE/CAPEC for weaknesses
and attack patterns), reducing ambiguity.
The eight integration rules (IR1--IR8) are declarative: each specifies a boolean criterion (Table~\ref{tab:rules}) that must be satisfied for a relation to be created. For instance, IR2 creates an \emph{enables} link between a surface and an attack only if at least one publication documents the corresponding entry vector. Once the extracted entities and evidence are fixed, the declarative rules deterministically produce the corresponding set of links.
Repeatability follows from the same explicit process: the construction and stopping conditions of Section~\ref{sec:construction} make the derivation reproducible.
This determinism concerns the \emph{application} of the rules to a given corpus; its relation to the completeness of that corpus is discussed in Section~\ref{sec:limitations}.

\smallskip\noindent\textbf{Usefulness and Navigability.}
Usefulness and navigability are supported by the interactive web visualizer
(Section~\ref{sec:visualizer}), which turns the catalog into navigable,
filterable views. Stakeholders can explore causal chains, filter risks by
deployment context, or project the catalog onto their role, reaching relevant
information without traversing the full catalog. Empirical validation with
practitioners is left to future work (Section~\ref{sec:agenda}).

\smallskip\noindent\textbf{Limitations of extraction.}
Entity and relation extraction may introduce subjectivity due to the interpretative nature of the process. We mitigate this through:
(i)~a systematic grounding in CWE/CAPEC identifiers;
(ii)~explicit integration rules (IR1--IR8); and
(iii)~four structural constraints verified programmatically: every attack must be linked to at least one surface~(C1), one weakness~(C2), and one risk~(C3), and every risk must affect at least one CIA dimension~(C4). These constraints are checked automatically during catalog generation, rejecting any entry that violates the meta-model schema.

\subsection{Cross-Cutting Observations}
\label{sec:observations}

By aggregating 43~publications into a single connected
structure, the catalog reveals patterns that
individual studies cannot capture.

\smallskip\noindent\textbf{Offensive/defensive imbalance.}
Of 43~publications, 31 are attack papers~(72\%) versus
9 defense papers~(21\%).
Many mitigations originate from the ``defenses'' section of
attack papers rather than from systematically evaluated defense
work, and several have already been bypassed: perplexity-based
detectors~\cite{greshake2023injection} by adaptive
attacks~\cite{liang2025saferag}, and similarity filtering by
reformulated queries~\cite{dimaio2024pirates}.
Bodea et~al.~\cite{bodea2026sok_rag_privacy} independently
report a similar 56\%/36\% attack-to-defense imbalance in
a privacy-focused RAG corpus of 72~publications.

\smallskip\noindent\textbf{Ingestion as structural hotspot.}
Surface S-DOC concentrates 16 of 38~attacks~(42\%), 14 of them
poisoning attacks.
Unlike transient injection attempts, poisoning can persist in
the corpus until detected or removed, making ingestion control
critical.
With trusted internal sources (P2\,=\,INTERNAL), four poisoning
risks are contextually mitigated; with open ingestion
(P2\,=\,OPEN), none of the 16~poisoning attacks is mitigated by
the ingestion property alone, and their applicability still
depends on the other deployment properties.
This leaves open-ingestion configurations particularly exposed
to integrity threats.

\smallskip\noindent\textbf{Output under-protection and coverage
gaps.}
Table~\ref{tab:assets} shows that OUTPUT (the answer returned
by the RAG system) has the highest gap ratio~(25\%):
R34 (inter-context conflict), R35 (soft ad injection), and
R36 (white DoS), all common to all RAG types, have no incoming
\emph{mitigates} edge (Table~\ref{tab:gaps}).
Two gaps also affect CORPUS: R17 (graph poisoning,
GRAPH-specific) and R33 (silver noise injection, common to all
RAG types).
This may reflect a publication bias or limitations of the
available corpus: output attacks are perceived as less novel or
impactful than poisoning attacks and receive less research
attention.

\begin{table}[t]
\caption{Per-asset gap analysis.}
\label{tab:assets}
\centering\scriptsize
\begin{tabular}{@{}lccc@{}}
\toprule
\textbf{Asset} & \textbf{Risks} & \textbf{Gaps} & \textbf{Gap ratio} \\
\midrule
CORPUS     & 20 & 2 (R17,R33)      & 10\% \\
OUTPUT     & 12 & 3 (R34,R35,R36)  & \textbf{25\%} \\
MEMBERSHIP & 5  & 0                & 0\% \\
EMBEDDINGS & 3  & 0                & 0\% \\
PROMPTS    & 1  & 0                & 0\% \\
\bottomrule
\end{tabular}
\end{table}

\subsection{Practitioner Use Cases}
\label{sec:implications}

Each view supports a distinct decision at a specific stage of
deployment.
\emph{At design time}, architects (V1) compare deployment
configurations (P1--P6) and assess risks before implementation;
for example, internal ingestion contextually mitigates four
poisoning risks.
\emph{In operation}, CISOs (V2) prioritize applicable risks and
identify coverage gaps.
\emph{During security assessment}, pentesters (V3) derive
configuration-specific test scopes from reachable attack chains;
an internal, isolated deployment eliminates two risks and
contextually mitigates four poisoning risks.
\emph{Across these stages}, DPOs (V4)
obtain per-asset information supporting GDPR and NIS2
assessments, including information relevant to
cross-border data transfers (GDPR Art.~46).

The web visualizer supports these use cases by projecting views,
filtering risks by configuration, and exposing causal chains.
It also allows new attacks to be drafted and checked against the
integration rules for inclusion in the catalog.
Whether these views effectively support practitioners remains an
empirical question, left for future work in
Section~\ref{sec:agenda}.

\subsection{Limitations and Threats to Validity}
\label{sec:limitations}

\smallskip\noindent\textbf{Catalog limitations.}
The catalog reflects the literature available and may require periodic updates; its extensibility
(IR2--IR4, IR7) allows new threats to be integrated without
structural change.
CWE coverage remains coarse: 14 of the 16~poisoning attacks map
solely to CWE-345, reflecting CWE's limited coverage of
ML-specific integrity flaws rather than a limitation of our
meta-model.

\smallskip\noindent\textbf{Internal rigor and completeness.}
The integration rules ensure consistent extraction: given the
same corpus, the same causal chains are derived.
They do not, however, guarantee completeness.
Publication bias, coarse CWE coverage, and genuinely unknown
threats limit what a literature-based approach can capture.
Completeness therefore holds only with respect to the
documented threat landscape, a blind spot the catalog makes
explicit through R0.

\smallskip\noindent\textbf{RAG specialization.}
RAG is deliberately scoped because deployment-aware filtering
and surface-level causal chains require a concrete
architecture.
The catalog does not transfer directly to other LLM
applications, but the meta-model structure and construction
process are architecture-agnostic and could be instantiated
for, e.g., agentic pipelines.

\smallskip\noindent\textbf{Methodological limitations.}
Attacks are modeled individually, so attack combinations
(e.g., chaining an injection with an exfiltration) are not
represented, although real-world adversaries may exploit such
sequences.

\smallskip\noindent\textbf{Threats to validity.}
\emph{Internal}: entity and relation extraction involves manual
judgment by a single coder, mitigated by systematic CWE/CAPEC
grounding and the boolean criteria of
Section~\ref{sec:validation}; inter-coder agreement was not
formally measured.
\emph{External}: the meta-model covers the three RAG types but has
not yet been validated on production deployments, and the four
views await practitioner evaluation (Section~\ref{sec:agenda}).
\emph{Construct}: definitions are grounded in ISO~27005, CWE,
and CAPEC, with the attack/risk distinction applied
consistently.

\subsection{Perspectives}
\label{sec:agenda}

The previous observations point to three priority directions.
1)~Closing the persistent coverage gaps: graph poisoning (R17) has no documented mitigation despite the rapid adoption of GraphRAG~\cite{gao2024rag_survey}; silver noise (R33) targets the CORPUS asset shared by all types; and the output-integrity cluster (R34--R36) combines a 25\% coverage gap with a probable publication bias.
2)~Re-evaluating mitigations proposed in attack papers, many of which lack systematic validation against adaptive adversaries.
3)~Empirical validation: a task-based user study in which practitioners of each role answer their Section~\ref{sec:implications} question on a real deployment.

\section{Conclusion}
\label{sec:conclusion}

We have presented a structurally specified security meta-model for RAG systems that, unlike existing frameworks, integrates an explicit causal chain from architecture to CIA impact, bridges RAG-specific threats to standardized weakness and attack-pattern identifiers (CWE/CAPEC), and adapts to deployment context through property-based risk filtering. Eight declarative integration rules verify and constrain how data from the literature is incorporated into a RAG security catalog.

Its practical value lies in three mechanisms: the filtering algorithm turns a large catalog into a deployment-specific risk profile; the four taxonomic views let each stakeholder reach the information relevant to their role; and the structural constraints, reinforced by the integration rules, enforce structural consistency as the catalog grows (with the limits discussed in Section~\ref{sec:limitations}). The cross-cutting analysis further reveals actionable patterns, notably the persistent under-protection of output integrity and the need to re-evaluate proposed defenses.

Three directions extend this work (Section~\ref{sec:agenda}): closing the five persistent coverage gaps, re-evaluating mitigations whose effectiveness remains unvalidated, and empirically validating the meta-model on production deployments with practitioner evaluation of the views.

\smallskip\noindent\textbf{Ethical statement.}
This work systematizes \emph{publicly documented} attacks; no new exploit is developed.

\smallskip\noindent\textbf{Open science.} Supplementary material (\url{https://anonymous.4open.science/r/rag-mmsecu/}) includes the full CSV database, regeneration scripts, the pregenerated Neo4j database, and the self-contained web visualizer supporting role-based views, causal-chain exploration, deployment filtering, and IR-validated attack drafting. Long-term Zenodo storage will be registered upon acceptance.

\ifdefined\limosversion
\smallskip\noindent\textbf{Acknowledgement.}
This work was supported by the Industrial Chair on Reliable and Confident Use of
LLMs\footnote{\url{https://uca-fondation.fr/chaires/chaire-sur-lutilisation-fiable-et-de-confiance-des-grands-modeles-de-langage-llms/}}
and by the MIAI Cluster, France 2030 (ANR-23-IACL-0006).
\fi

\bibliographystyle{splncs04}
\label{refs-start}
\bibliography{refs}

\appendix

\begin{center}
\textbf{Supplementary Material}
\end{center}

\noindent This supplementary material provides a full-catalog
overview (Appendix~\ref{app:overview}) and additional tables
referenced from the discussion (Appendix~\ref{app:tables}).
It is optional: reviewers are not required to read it.

\section{Full Catalog Overview}
\label{app:overview}

\begin{sidewaysfigure}
\centering
\includegraphics[width=\textheight,height=.94\textwidth,keepaspectratio]{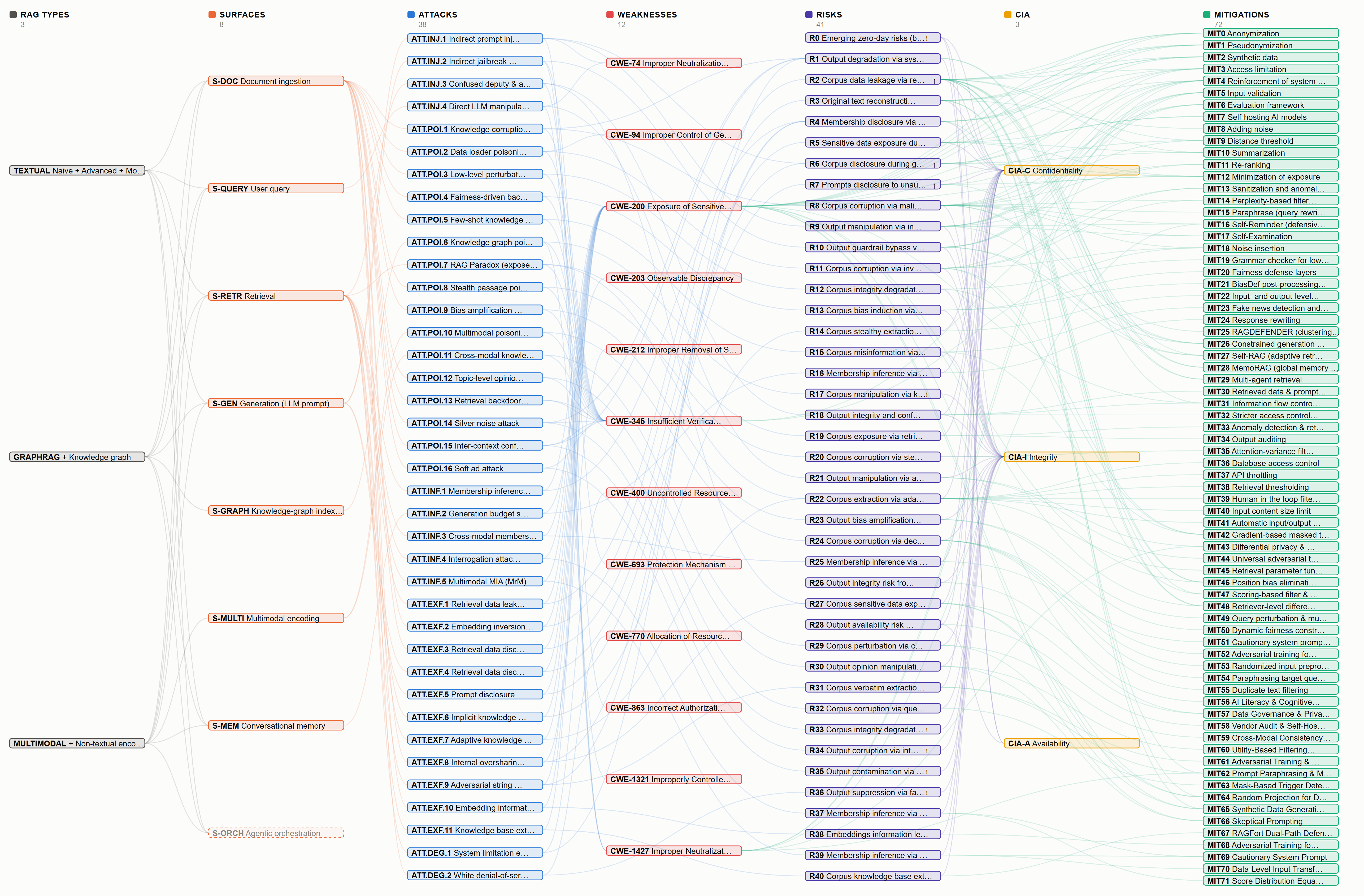}
\caption{The complete catalog (177~nodes), as rendered by
the web visualizer under the default deployment configuration. Columns
follow the causal chain RAG type $\rightarrow$ Surface
$\rightarrow$ Attack $\rightarrow$ Weakness / Risk $\rightarrow$ CIA,
with Mitigations linked to the risks they mitigate. Markers:
``!''\,=\,coverage gap (risk without mitigation);
``$\uparrow$''\,=\,risk aggravated by the default configuration;
dashed\,=\,not applicable to it.}
\label{fig:overview}
\end{sidewaysfigure}

Figure~\ref{fig:overview} renders the complete catalog in
one view: the 3~RAG types, 8~surfaces, 38~attacks, 12~weaknesses,
41~risks, 3~CIA dimensions and 72~mitigations with all their
relations. Long node names are truncated to fit; full names,
descriptions and sources are available interactively in the web
visualizer and as zoomable vector exports in the supplementary
material (\url{https://anonymous.4open.science/r/rag-mmsecu/}).

\clearpage
\section{Additional Tables}
\label{app:tables}

\begin{table}[!ht]
\caption{Discriminant power of candidate segmentation criteria.}
\label{tab:discrim}
\centering\scriptsize
\begin{tabular}{@{}lcccc@{}}
\toprule
\textbf{Criterion} & \textbf{Total} & \textbf{Shared} &
  \textbf{Type-specific} & \textbf{Power} \\
\midrule
Weakness (CWE) & 12 & 12 & 0 & None \\
Surface         &  8 &  6 & 2 & Low \\
Attack          & 38 & 33 & 5 & Medium \\
\textbf{Risk}   & 41 & 36 & 5 & Medium+CIA \\
\bottomrule
\end{tabular}
\end{table}

\begin{table}[!ht]
\caption{Persistent coverage gaps.}
\label{tab:gaps}
\centering\scriptsize
\begin{tabular}{@{}llll@{}}
\toprule
\textbf{Risk} & \textbf{Category} & \textbf{Type} & \textbf{Explanation} \\
\midrule
R17 (Graph poison.)     & POI & GRAPH only & No triplet validation \\
R33 (Silver noise)      & POI & ALL        & No noise filtering \\
R34 (Inter-context)     & POI & ALL        & No contradiction detection \\
R35 (Soft ad)           & POI & ALL        & No semantic filtering \\
R36 (White DoS)         & DEG & ALL        & No defense \\
\bottomrule
\end{tabular}
\end{table}

\end{document}